# An optimised cuckoo-based discrete symbiotic organisms search strategy for tasks scheduling in cloud computing environment

Suleiman Sa'ad[a,b,1,*], Abdullah Muhammed[a,1,*], Mohammed Abdullahi[c,1], Azizol Abdullah[a]

[a] Department of Communication Technology and Networks, Universiti Putra Malaysia, 43400 UPM, Serdang Selangor, Malaysia

[b] Department of Information Technology, Modibbo Adama University of Technology Yola 640231, Yola Nigeria

[c] Department of Computer Science, Ahmadu Bello University Zaria 810107, Zaria Nigeria

[*] Corresponding authors.

*E-mail addresses:* suleimanu@mau.edu.ng (Suleiman Sa'ad), abdullah@upm.edu.my (Abdullah Muhammed).

[1] These authors contributed equally to this work.

## Abstract

Currently, the cloud computing paradigm is experiencing rapid growth as there is a shift from other distributed computing methods and traditional IT infrastructure towards it. Consequently, optimised task scheduling techniques have become crucial in managing the expanding cloud computing environment. In cloud computing, numerous tasks need to be scheduled on a limited number of diverse virtual machines to minimise the imbalance between the local and global search space; and optimise system utilisation. Task scheduling is a challenging problem known as NP-complete, which means that there is no exact solution, and we can only achieve near-optimal results, particularly when using large-scale tasks in the context of cloud computing. This paper proposes an optimised strategy, Cuckoo-based Discrete Symbiotic Organisms Search (C-DSOS) that incorporated with Levy-Flight for optimal task scheduling in the cloud computing environment to minimise degree of imbalance. The strategy is based on the Standard Symbiotic Organism Search (SOS), which is a nature-inspired metaheuristic optimisation algorithm designed for numerical optimisation problems. SOS simulates the symbiotic relationships observed in ecosystems, such as mutualism, commensalism, and parasitism. To evaluate the proposed technique, the CloudSim toolkit simulator was used to conduct experiments. The results demonstrated that C-DSOS outperforms the Simulated Annealing Symbiotic Organism Search (SASOS) algorithm, which is a benchmarked algorithm commonly used in task scheduling problems. C-DSOS exhibits a favourable convergence rate, especially when using larger search spaces, making it suitable for task scheduling problems in the cloud. For the analysis, a t-test was employed, reveals that C-DSOS is statistically significant compared to the benchmarked SASOS algorithm, particularly for scenarios involving a large search space.

# 1. Introduction

Cloud computing is one of the recent developments in the field of computing which enables limitless usage of Information Technology (IT) resources in diverse domains such as education, medicine, business, mobile system, smart systems, environmental computing etc. [1], [2]. This has led to rapid adoption of cloud computing in recent years, because it acts as an efficient computing paradigm for renting IT services and infrastructures based on pay-per-use model [3]. The model eliminates the need for companies to invest in acquisition of IT infrastructures or software licenses.

Cloud services are categorized as Software as a Service (SaaS), Platform as a Service (PaaS), and Infrastructure as a Service (IaaS) [4], [5]. These services are provisioned to users of virtual resources through virtualization and containerisation, which make cloud computing resources dynamic and elastic thereby creating the notion of unlimited resources. Users are charged for the services they consumed on pay-per-use basis, and this flexible mode of charging users has encouraged migration of IT services to the cloud environment [5]. Therefore, the focus of this paper is on IaaS cloud where computing resource are offered as services. Users subscribed for VMs for execution of their tasks, and better utilisation of physical resources is directly dependent on the optimal scheduling of tasks on the VMs.

Task scheduling has been one of the widely researched problems in cloud computing, but it remains an NP-hard problem [6]. Pool of Virtual resources are made available to cloud users by network of servers in IaaS layer. IaaS layer delivers hardware and associated software, which enable provision of flexible and efficient computational capacities to end users. The resource management subsystem of IaaS layer is responsible for scheduling submitted tasks for execution. Scheduling of tasks on VMs is a key process on IaaS cloud, because mapping of tasks to VMs need to be carried out in an efficient manner due to heterogeneous and dynamic characteristics of VMs. Since there is no exact algorithm for finding optimal solution for NP-Complete problems, a good scheduling solution can only be achieved via heuristic methods [7]–[10]. The objective of task scheduling algorithm is to reduce certain parameters such as, degree of imbalance, makespan, execution time and cost; the algorithm decides which VM should execute the received task. In cloud computing environment, VMs have heterogeneous processing capacities and characteristics. Therefore, load balancing among VMs needs to be taken into account when scheduling tasks, which entails careful coordination and optimisation in order to achieve lower makespan [11]–[13]. Task scheduling algorithms try to efficiently balance the load of the system taking into consideration total execution time of available VMs.

Methods proposed in the literature for solving task scheduling problems are either heuristic or metaheuristic based. Heuristic based methods try to find optimal solution based on some predefined rules, and the quality of solutions obtained by these methods are dependent on the underlining rules and problem size. The solution obtained by heuristics search methods are not feasible and they are generated at high operating cost [14], [15].

Therefore, metaheuristic algorithms have been extensively applied to solve optimisation problems. Metaheuristic methods employ a pool of candidate solutions to traverse solution space unlike the mathematical and heuristic techniques that uses single candidate solution. This attribute of metaheuristic algorithms make them perform better than mathematical and heuristic strategies. Some of the popular metaheuristic methods for solving task scheduling problems in cloud computing environment include Genetic Algorithm [16] and

Particle Swarm Optimisation [17]. Furthermore, Ant Colony Optimisation [18], [19], League Championship Algorithm [11], BAT algorithm [20], [21], Symbiotic Organisms Search [12], [13]. The idea of SOS as a metaheuristic algorithm was introduced in [22]. SOS algorithm was inspired by interactive relationship exhibited by organisms in ecosystem for survival and it was shown to perform competitively well [22] with Genetic Algorithm (GA), Differential Evolution (DE), Particle Swarm Optimisation (PSO), Honey Bee Colony (HBC) etc. Since the introduction of SOS algorithm, a number of researches have applied SOS to solve some practical optimisation problems [12], [23]–[27]. Hence, the potential of SOS in finding global solution to optimisation problems exhibited so far make it attractive for further investigation and exploration.

Furthermore, quality of solution and convergence speed obtained by metaheuristic algorithms can be improved by its hybridization with either a metaheuristic algorithm or heuristic strategy, by generating initial solution using heuristic search techniques or by modifying the transition operator [28], [29]. To the best of our knowledge, none of the aforementioned strategies has been explored to investigate the possible improvement of SOS in terms of convergence speed and quality of solution obtained by SOS.

Therefore, in this paper, a fitness function model for computing degree of imbalance was developed, taking into account utilisation of VMs to minimise degree of imbalance between the global and local search space thereby optimising the VM's utilisation. Task scheduling strategies were studied using Improved Symbiotic Organism Search (SASOS). The proposed Cuckoo-based Discrete Symbiotic Organism Search (C-DSOS) strategy combines Cuckoo search (CS) method and SOS variant by [30]. The SOS uses fewer control parameters, and has a strong exploration as well as faster convergence capability. CS was used to search local solution space identified by SOS, which equip C-DSOS with exploitative ability. The objective is to obtain optimal schedules by minimising degree of imbalance within the search space.

The main contributions of the paper are:

i. An objective function for optimum scheduling of tasks on VMs is presented taking into account the utilisation level of VMs in order to minimise degree of imbalance within the search space.

ii. Hybridization of SOS with Cuckoo search levy flight section, applying the levy flight to find optimum solution in the global solution regions identified by DSOS.

iii. Implementation of the proposed method in CloudSim.

iv. Performance comparison of SASOS and the proposed method in terms of response time and degree of imbalance.

v. Empirical analysis of convergence speed obtained by C-DSOS and SASOS.

The organisation of remaining parts of the paper is as follows. Related work on metaheuristic algorithms applied to task scheduling problems in cloud and SOS are presented in Section 2. Design of the proposed algorithm and its description is presented in Section 3. Section 4 describes problem formulation. Results of simulation and its discussion are in Section 5. Section 6 presented conclusion of the paper.

## 2. Related works

Metaheuristic algorithms such as [11], [13], [27], [31]–[39] have been applied to solve task assignment problems in order to reduce makespan and response time. These methods have proven to find optimum mapping of workloads to resources, which reduces cost of computation, better quality of service, and increased utilisation of computing resources. ACO, GA, and their variants are the mostly used nature inspired population based algorithms in the cloud. However, PSO consistently excels over GA and ACO across various scenarios [40] and has faster execution time. Implementing PSO is notably more straightforward compared to GA and ACO. respectively. The application of PSO in addressing workflow scheduling problems has been extensively explored in previous studies [17], [41]–[43] with the overarching goal of mitigating communication costs and minimising makespan. The scheduling of independent tasks in cloud computing has been a subject of investigation utilising PSO, as evidenced by studies of [44]–[46]. These strategies has demonstrated its effectiveness in achieving minimal makespan. Furthermore, enhanced and hybrid iterations of PSO, proposed by [44], [45], [47], have yielded superior results for task scheduling in the cloud when compared to ACO and GA. Recently, discrete versions of the SOS methods were applied to task scheduling problems within the cloud computing environment [12], [13], [27], [37]. These applications revealed that the SOS algorithm surpassed both nature-inspired metaheuristics and their popular variants in terms of performance.

### 2.1 Symbiotic organisms search

The SOS algorithm draws inspiration from the symbiotic interactions observed between paired organisms within an ecosystem. In this context, each organism symbolises a potential solution to an optimisation problem, with its unique position within the solution space. These organisms adapt their positions based on biological interaction models such as mutualism, commensalism, and parasitism, reminiscent of the ecological dynamics. In the initial phase of the algorithm, the concept of mutualism comes into play, where both interacting organisms derive benefits from their relationship. Subsequently, in the second phase of the algorithm, commensalism is applied; a form of interaction where one organism benefits while the other remains unaffected. This phase serves to refine the solution space.

Furthermore, the third phase introduces parasitism interaction, where only one organism benefits at the expense of the other. This interaction technique is implemented to further optimise the solution space. Through these interactions, the fittest organisms thrive in the solution space, while the less fit ones are gradually eliminated. The best-performing organisms are identified based on their cumulative benefits from all three phases of interaction. These phases are iteratively applied to a population of organisms representing candidate solutions until the predefined stopping criteria are met. The quality of an organism's position is contingent upon its fitness, which quantifies the extent to which the organism has adapted to the ecosystem. For a detailed outline of the SOS algorithm, refer to Algorithm 1.

```
Algorithm 1: Standard - Symbiotic Organisms Search
INPUT:
  - Ecosize (Size of the ecosystem)
  - Initial population (Initial candidate solutions)
  - Stopping criteria (Condition to stop the algorithm)
OUTPUT:
  - Optimal solution
1: Initialize the ecosystem with the given Initial population
2: Initialize a variable to keep track of the number of iterations (iteration_count)
3: Repeat the following steps until the stopping condition is met:
4:    For each organism in the ecosystem:
```

```
5:      Determine the best organism in the current ecosystem
6:      Apply the Mutualism Phase to the organism
7:      Apply the Commensalism Phase to the organism
8:      Apply the Parasitism Phase to the organism
9:    End For
10:   Update the iteration_count by 1
11: Until stopping condition is exceeded (e.g., a maximum number of iterations is reached)
12: Return the best organism found as the Optimal solution
```

Moreover, SOS shares several common features with other nature-inspired algorithms. It employs a population of organisms to represent candidate solutions by utilising operators to guide the search process. A selection mechanism is employed to retain superior solutions, necessitating the pre-definition of population size and stopping criteria before initiating the search process. In contrast to algorithms like PSO, which rely on specific parameters such as inertia weight, social, and cognitive factors, or GA that utilise crossover and mutation, SOS does not demand algorithm-specific parameters. The improper tuning of these parameters in other algorithms can potentially lead to suboptimal solutions.

The SOS algorithm commences with the creation of a randomly generated population of organisms, referred to as an ecosystem. Subsequently, the positions of these organisms are updated using the three distinct phases of the SOS. In a D-dimensional solution search space, a population of n organisms is denoted as $X = \{X_1, X_2, X_3,...,X_n\}$, with each *ith* organism's position represented as $X_i = \{X_{i1}, X_{i2}, X_{i3}, ..., X_{id}\}$. Furthermore, to evaluate the quality of a solution obtained by an organism, a fitness function is defined. Each organism in the population signifies a task schedule, encoded in a vector of dimension *1xn*, where *n* represents the number of tasks. The elements of this vector are natural numbers within the range *[0, m − 1]*, where *m* denotes the number of available resources for task execution.

The best searched position by all organisms up to the current point is denoted as $X^{best}$. Given that task scheduling is a discrete optimisation problem, and SOS was originally designed for continuous optimisation, in the works of [12], [13], [37] adapt a mapping function to translate continuous positions into discrete positions. This mapping function is defined in Equation (1). The fitness value of each organism is assessed iteratively, utilising their corresponding positions as input in the three phases of the algorithm, as elucidated in the subsequent subsections.

$$X_i = round(X_i \bmod m) \tag{1}$$

Here, *m* represents the number of available resources for task execution.

Mutualism Phase: in this phase, we let $X_i$ represent the *ith* individual in the ecosystem. In this phase, $X_j$ is randomly chosen from the swarm of organisms to engage in an interaction with $X_i$, where $i \neq j$, and the goal is mutual benefit. The purpose of this interaction is to enhance the survival prospects of both $X_i$ and $X_j$ within the ecosystem. The updated candidate solutions for $X_i$ and $X_j$ are derived in accordance with Equations (2) and (3) [30].

$$X_i^* = X_i + U(0,1) * (X^{best} + MV * \alpha) \tag{2}$$

$$X_j^* = X_j + U(0,1) * (X^{best} + MV * \beta) \tag{3}$$

$$MV = \frac{1}{2}(X_i + X_j) \qquad (4)$$

Where *i* takes on values from *1* to *ecosize* and *j* spans from *1* to *ecosize*, excluding the case where j is equal to i. The symbol *U(0, 1)* represents a vector of uniformly distributed random numbers ranging between *0* and *1*. *MV* denotes the mutual relationship vector between $X_i$ and $X_j$, defined in Equation (4). $X^{best}$ signifies the organism with the best fitness value. *α* and *β* stand for the benefit factors governing the interaction between organism $X_i$ and $X_j$. Furthermore, within a mutual relationship, an organism can experience varying degrees of benefit while interacting with a mutual partner. Therefore, *α* and *β* are probabilistically determined, taking on values of either *1* or *2*, where *1* denotes a lighter benefit, and *2* represents a heavier benefit.

Consequently, the new candidate solutions replace the old ones if their fitness values prove superior. In such cases, $X^*_i$ and $X^*_j$ take the place of $X_i$ and $X_j$, respectively, in the subsequent generation of the ecosystem. Conversely, if the fitness values of $X^*_i$ and $X^*_j$ are not better, they are discarded, while $X_i$ and $X_j$ continue on to the next generation of the ecosystem. Equations (5) and (6) capture this specific scenario [30].

$$X_i = \{X^*_i \text{ if } f(X^*_i) > f(X_i) \text{ or } (X_i \text{ if } f(X^*_i) \le f(X_i)\} \qquad (5)$$

$$X_j = \{X^*_j \text{ if } f(X^*_j) > f(X_j) \text{ or } (X_j \text{ if } f(X^*_j) \le f(X_j)\} \qquad (6)$$

Here, *f(.)* represents the fitness evaluation function.

Commensalism Phase: Within the commensalism phase, an ecosystem member represented as *ith* engages in a random selection process to interact with another organism, denoted as $X_j$, where *i* does not equal *j*. In this scenario, $X_i$ seeks to derive a benefit from its interaction with $X_j$, while $X_j$ may experience either gain or loss because of this interaction. Equation 7 captures the mathematical representation of this interaction [30].

$$X^*_i = X_i + U(-1,1) * (X^{best} + X_j) \qquad (7)$$

In this context, *U(-1, 1)* denotes a vector comprising uniformly distributed random numbers ranging from *-1* to *1*. Similar to the mutualism phase, $X^{best}$ designates the organism with the highest fitness value. If the fitness value, denoted as $f(X^*_i)$, surpasses the current fitness value of $X_i$, represented as $f(X_i)$, then $X_i$ is updated to $X^*_i$ as calculated by Equation 7. The mechanism for updating $X_i$ is defined by Equation 8 [30].

$$X_i = (X^*_i \text{ if } f(X^*_i) > f(X_i) \text{ or } (X_i \text{ if } f(X^*_i) \le f(X_i) \qquad (8)$$

Parasitism Phase: Within the parasitism phase, an artificial parasite referred to as a "parasite vector" is generated by cloning the *ith* organism, $X_i$, and subjecting it to alterations through a randomly generated number. Subsequently, a random selection is made from the ecosystem, resulting in the choice of $X_j$, and the fitness values of both the parasite vector and $X_j$ are computed. If the parasite vector demonstrates superior fitness compared to the $X_j$, the parasite vector supplants $X_j$. Conversely, if $X_j$ exhibits better fitness, it endures into the next generation of the ecosystem, and the parasite vector is discarded. The update of $X_j$ adheres to the relationship outlined in Equation 9 [30].

$$X_j = (PV \text{ if } f(PV) > f(X_j) \text{ or } (X_j \text{ if } f(PV) \le f(X_j) \qquad (9)$$

Where, PV stands for the parasite vector.

## 2.2 Cuckoo search algorithm

The Cuckoo Search (CS) algorithm is a metaheuristic algorithm inspired by the remarkable breeding behaviour of Cuckoo birds, known for their highly aggressive reproduction strategies. CS incorporates a key element called the Levy flight, which involves an initial exploratory phase in the search space followed by an exploitation phase, effectively harmonizing the search process between local and global exploration. As a result, CS maintains a remarkable equilibrium within the search space, balancing the quest for diverse solutions with the focus on exploiting the most promising ones. Equation 10, as presented by [48], describes the Levy Flight component in detail.

$$X_i^t = X_i^{t+1} + \alpha \; \Theta Levy(\lambda) \tag{10}$$

Where $X_i^t$ is the new solution, $X_i^{t+1}$ is the current solution and $\alpha \; \Theta \; Levy(\lambda)$ is the transaction probability.

## 3. Cuckoo-based discrete symbiotic organism search (C-DSOS) strategy

The basic Symbiotic Organism Search algorithm (SOS) encompasses several key features, including parameterless control and a structured sequence of three phases: mutualism, commensalism, and parasitism. The absence of control parameter fine-tuning, due to its parameterless nature, simplifies its implementation. The mutualism and commensalism phases focus on local search aspects, while the parasitism phase addresses the global search dimension of the algorithm. Within the parasitism phase, a random organism $(X_k)$ from the ecosystem replaces the global best candidate $(X^{best})$ to introduce diversity into the solution space. Additionally, the inclusion of Levy Flight within the mutualism phase enhances the algorithm's performance by accelerating convergence, preventing local optima traps, and mitigating imbalance issues encountered in load balancing problems, as exemplified in Equations 11 to 14.

$$S_1(p) = X_i + (eq.10) * r'\{X_k - (X_i + X_j/2)f_1\} \tag{11}$$

$$S_2(p) = X_i + (eq.10) * r''\{X_k - (X_i + X_j/2)f_2\} \tag{12}$$

$$X_I^*(q) = \lceil S_1(p) \rceil \; mod \; m + 1 \tag{13}$$

$$X_j^*(q) = \lceil S_2(p) \rceil \; mod \; m + 1 \tag{14}$$

Where, $X_I^*(q) \; and \; X_j^*(q)$ are the modified positions of the *ith* and *jith* members of the ecosystem with $i \neq j$, $X_j$ is the organism randomly selected in the ith iteration also for $i \neq j$, $f_1$, $f_2 \in 1,2$ are randomly determined and represent the benefit factor from mutual interaction, where $r', r'' \in rand(0,1)$ are random numbers generated uniformly in the specified interval and ⌈⌉ is an upper bound function.

Furthermore, equations 15 to 19 illustrate that the commensalism phase, working in conjunction with the mutualism phase, executes local search operations and narrows the range of random values from *[-1, 1]* to *[0.4, 0.9]* (Do et al., 2018) to further enhance convergence speed and reduce imbalance.

$$r' \leftarrow rand[0.4, 0.9] \tag{15}$$

$$S_3(p) = X_i + r'(X^{best} - X_j) \tag{16}$$

$$X_i^*(q) = \lceil S_3(p) \rceil \bmod m + 1 \tag{17}$$

$$S_4(p) = r_4 X_i \tag{18}$$

$$X^p(q) = \lceil S_4(p) \rceil \bmod m + 1 \tag{19}$$

Where $r'$ in equation 15 as in the work of [49] is a random number uniformly generated between 0.4 and 0.9.

Despite the advantages of local search, it is essential to acknowledge that it may not always yield superior results due to the risk of premature convergence leading to local optima traps. This issue impacts task scheduling efforts aimed at minimising imbalance. Given SOS's efficacy in addressing NP-hard problems, as documented in previous literature, modifications are introduced in both the mutualism and commensalism phases. Specifically, Levy Flight is integrated into the local search aspect of the discrete symbiotic organisms search strategy. Consequently, the CDSOS strategy attains a balanced exploration of both local and global solution spaces, yielding improved outcomes. This adaptation results in reduced computational time and, consequently, minimises task scheduling imbalances. Consequently, this study appropriately applies the CDSOS task scheduling strategy to determine the allocation of tasks to virtual machines.

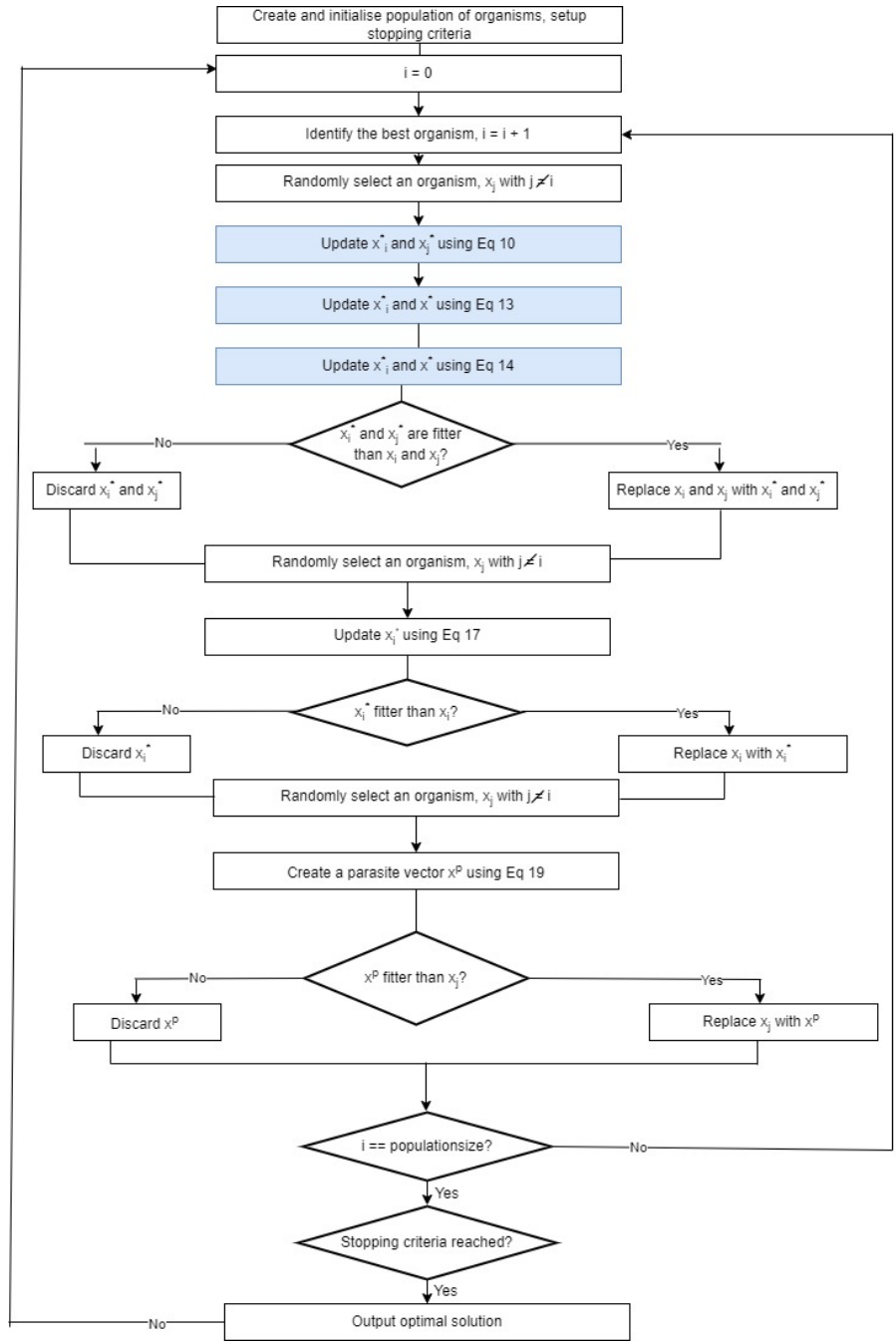

**Figure 1: The flowchart of the proposed CDSOS strategy**

Therefore, the hybrid strategy is intended to minimise the degree of imbalance in the search space, thereby increasing its convergence rate and balancing the load. Therefore, it is important that the search results are returned more efficiently due to the improved search efficiency of the proposed strategy. The flowchart of the proposed CDSOS task scheduling strategy is shown in Figure 1. The Figure shows the scheduling process that generates a task scheduling sequence that guarantees minimum degree pf imbalance in the later stage. Algorithm 2 illustrates the pseudo-code of the CDSOS task scheduling strategy.

**Algorithm 2** Cuckoo-based Discrete Symbiotic Organism Search Algorithm
1: Create and Initialise the Population of Organisms in the Ecosystem $X = \{x_1, x_2, x_3...y_N\}$
2: Setup stopping criteria
3: $iteration\_number \leftarrow 0$
4: $x^{best} \leftarrow 0$
5: **Do**
6:   $iteration\_number \leftarrow iteration\_number + 1$
7:   $i \leftarrow 0$
8:   **Do**
9:     $i \leftarrow i + 1$
10:   **for** $j = 1$ $to$ $N$ **do**
11:     **if** $f(x_j) > f(x^{best})$ **then**
12:       $x^{best} \leftarrow y_j$
13:     **end if**
14:   **end for**
15:   //**Mutualism phase**
16:   Randomly select $x_j$ with $i \neq j$
17:   Update $x_i^*$ and $x_j^*$ using Equations 13 and 14
18:   **if** $f(x_i^*) > f(x_i)$ **then**
19:     $x_i \leftarrow x_i^*$
20:   **end if**
21:   **if** $f(x_j^*) > f(x_j)$ **then**
22:     $x_j \leftarrow x_j^*$
23:   **end if**
24:   //**Commensalism phase**
25:   Randomly select $x_j$ with $i \neq j$
26:   $r_3 \leftarrow$ using Equation 15
27:   Update $x_i^*$ according Equation 17
28:   **if** $f(x_i^*) > f(x_i)$ **then**
29:     $x_i \leftarrow x_i^*$
30:   **end if**
31:   //**Parasitism phase**
32:   Randomly select $x_j$ with $i \neq j$
33:   Create a parasite vector $x^p$ from $x_i$ using random number according Equation 19
34:   **if** $f(x^p) > f(x_j)$ **then**
35:     $x_j \leftarrow x^p$
36:   **end if**
37: **While** $i <= N$           2
38: **While** stopping condition is not true

## 4. Problem formulation

The task is formulated to align with the objective function, ensuring that each task design solution is assessed in a fair manner with respect to meeting the general requirements. This assessment is facilitated through the utilisation of the objective function. Specifically, the objective function plays a crucial role in obtaining a polynomial time approximation within the framework of the proposed CDSOS strategy. Its purpose is to generate results that closely strategy the optimal value, effectively characterising how well the proposed strategy achieves the optimised degree of imbalance. In essence, scheduling entails the efficient allocation of tasks to selected virtual machines, aiming to minimise the degree of imbalance. This objective is intricately linked to the main goal of task scheduling.

Consider a set of independent tasks to be scheduled for processing on a set of heterogeneous VMs. $V = \{v_k \mid m \geq k \geq 1\}$, are the sets of VMs, where $m$ is the number of VMs. $T = \{t_i \mid n \geq i \geq 1\}$, denotes the task groups and $n$ is the total number of tasks. The goal is to efficiently assign each task $t_i \; \forall \; i = \{1,2,\ldots,n\}$ to a matching cloud virtual machine $v_k \; \forall \; k = \{1,2,\ldots,m\}$ to minimise the degree of imbalance. Thus, the execution time of the task $t_i$ processed on a virtual machine $v_k$ is calculated using Equation 20.

$$exe_k = \sum x_{ik} * {t_{ik}}/{npe_k} \times v_{mipsk} \qquad (20)$$

$i \in Task, k \in virtual\ machine$

Where $exe_k$ represents the execution time of the task allocated on $v_k$; $x_{ik}$ is equal $1$ in the case that task $i$ is allocated on virtual machine $k$, and $zero(0)$; otherwise $t_{ik}$ is the amount of task allocated on $v_k$; $v_{mipsk}$ is the length of a task in Million Instructions (MIs); $v_{mipsk}$ is the speed of $v_k$ in Millions Intructions per Second (MIPS) and $npe$ denotes the number of processing elements. If more than, one virtual machine $v_k \; \forall \; k = 1,2,\ldots,m$ processes the task $t_i \; \forall \; k = 1,2,\ldots,n$, the total execution time of the task processed on all virtual machines $v_k$ is calculated using Equation 21.

$$T_{exek} = \sum exe_k \qquad (21)$$

$\forall i = \{1,2,\ldots,n\} \; k = \{1,2,\ldots,m\}$

Consequently, Equation 22 presents the objective function for the problem, aiming to minimise the degree of imbalance (DI).

$$DI = \frac{T_{max} - T_{min}}{T_{avg}} \qquad (22)$$

$\forall i = \{1,2,\ldots,n\} \; k = \{1,2,\ldots,m\}$

The DI equation, derived from the model introduced by [50], serves as a valuable tool for evaluating the effectiveness of any task scheduling solution that has been developed. It allows us to assess how well such a solution can address real NP-hard scheduling challenges. Furthermore, this model can be seamlessly incorporated into a nature-inspired metaheuristic algorithm to facilitate the quest for optimal solutions.

## 5. Simulation and results

The proposed task scheduling strategy is implemented using the CloudSim simulator, as introduced by [51]. To facilitate the simulation, the datacentre broker policy within CloudSim is extended. The choice of properties for virtual machines, hosts, and tasks closely follows the conventions utilised in numerous prior studies. These properties are outlined in Table 1.

**Table 1: Datacentre parameter settings**

| Entity Type | Parameter | Value |
|---|---|---|
| Datacentre | No. of datacentre | 2 |
| | No. of host | 2 |
| | Host RAM | 20 GB |
| | Storage | 1 TB |
| | Bandwidths | 10 GB/s |
| | Accumulated host processing power | 1000000 MIPS |
| Cloudlets | Lengths | 100 - 1000 MIs |
| | No. of Cloudlets (tasks) | 100 - 1000, NASA Ames iPSC/860 and HPC2N workloads |
| Virtual Machine (VM) | No. of VMs | 25 |
| | VMs Monitor Operation System RAM | Xen Linux |
| | Storage Bandwidth | 0.5 GB 10 GB 1 GB/s |
| | VMs processing power | 500 - 5000 MIPS |
| | Processing element | 1 |
| | VM Policy | Time-shared |

Within this context, special attention is given to the selection of virtual machine (VM) sizes, with an inquiry into whether the VM dimensions could influence the proposed task scheduling strategy. In line with standard practices in the field of task scheduling in cloud computing, its length characterises each task, and this length varies among tasks, rendering each one unique. Hosts are equipped with a storage capacity of one terabyte (1TB) to support VMs in each datacentre. Many previous studies in the domain of task scheduling in cloud computing, such as [12], [13], [30], [32], [37], [39], [52] have adopted these property settings to evaluate their task scheduling and resource provisioning techniques.

The inertia weight values and coefficient factors, presented in Table 2, are selected in accordance with their usage in previous studies, such as [29], [30], [53]–[55].

**Table 2: Parameter settings for the scheduling approaches** [56]

| Algorithm | Parameter | Value |
|---|---|---|
| Symbiotic Organisms Search (SOS) | Number of organisms | 100 |
| | Number of iterations | 1000 |
| Simulated Annealing (SA) | Initial temperature, *Finit* | 10 |
| | Final temperature, *Ffinal* | 0.001 |
| | Cooling rate, $\delta$ | 0.9 |

The task instances are generated from synthetic workloads, and for comparison purposes, the Simulated Annealing Symbiotic Organism Search (SASOS) algorithm, as proposed by [30], is employed. SASOS serves as a benchmarking metaheuristic algorithm to assess the improvements made in the mutualism and commensalism phases of the DSOS algorithm, particularly in terms of local search. SASOS is employed within the context of a cloud computing environment to address the heterogeneous task scheduling problem and optimise the degree of imbalance. Therefore, this paper has adapted the algorithm as an enhancement to the standard SOS algorithm, primarily to mitigate premature convergence and minimise the degree of imbalance. A comparative analysis between this algorithm and the proposed CDSOS strategy offers valuable insights into the efficiency of the latter.

Moreover, the results obtained from a simulation conducted by executing the proposed CDSOS approach alongside a comparative algorithm for 10,000 iterations illustrate the effectiveness of the proposed strategy, particularly in the context of heterogeneous virtual machines (VMs). This strategy sheds light on the significance of achieving varying degrees of imbalance. Consequently, the simulation results are presented by examining maximum, minimum, and average values. The comparison with SASOS primarily centres around evaluating the degree of imbalance. For an in-depth analysis of the results, please refer to Figures 2 to 5, while detailed simulation outcomes can be found in Tables 3 to 6.

**Table 3: Comparison of degree of imbalance obtained by SASOS and CDSOS for normal distribution dataset**

| Task Size | SASOS | | | CDSOS | | | | |
|---|---|---|---|---|---|---|---|---|
| | max | min | avg | max | min | avg | t-value | p-value |
| 100 | 18.59 | 16.55 | 17.41 | 18.59 | 15.77 | 17.63 | 1.6660 | 0.10110 |
| 200 | 19.44 | 18.53 | 18.75 | 19.44 | 17.32 | 18.54 | -1.8174 | 0.07433 |
| 300 | 19.44 | 19.05 | 19.19 | 19.44 | 16.24 | 18.49 | -4.6371 | 2.06E-05 |
| 400 | 19.65 | 19.31 | 19.38 | 19.65 | 17.02 | 18.80 | -4.8204 | 1.07E-05 |
| 500 | 19.67 | 19.42 | 19.52 | 19.67 | 18.04 | 19.13 | -4.6988 | 1.65E-05 |
| 600 | 19.75 | 19.54 | 19.59 | 19.75 | 17.86 | 19.23 | -4.5002 | 3.32E-05 |
| 700 | 19.80 | 19.61 | 19.66 | 19.80 | 18.80 | 19.37 | -5.5718 | 6.84E-07 |
| 800 | 19.91 | 19.62 | 19.7 | 19.91 | 18.77 | 19.46 | -4.7656 | 1.30E-05 |
| 900 | 19.91 | 19.70 | 19.74 | 19.91 | 18.40 | 19.45 | -4.8315 | 1.03E-05 |
| 1000 | 19.91 | 19.73 | 19.76 | 19.91 | 18.33 | 19.53 | -3.9145 | 0.00024 |

**Table 4: Comparison of degree of imbalance obtained by SASOS and CDSOS for left-half distribution dataset**

| Task Size | SASOS | | | CDSOS | | | | |
|---|---|---|---|---|---|---|---|---|
| | max | min | avg | max | min | avg | t-value | p-value |
| 100 | 19.91 | 16.99 | 17.35 | 19.91 | 16.50 | 17.98 | 6.3475 | 3.62E-08 |
| 200 | 19.91 | 18.26 | 18.77 | 19.91 | 17.49 | 18.66 | -1.2621 | 0.21200 |
| 300 | 19.91 | 19.03 | 19.15 | 19.91 | 16.99 | 18.75 | -3.0073 | 0.00389 |
| 400 | 19.91 | 19.26 | 19.38 | 19.91 | 16.66 | 18.75 | -4.5462 | 2.83E-05 |
| 500 | 19.91 | 19.47 | 19.52 | 19.91 | 17.10 | 19.15 | -3.3504 | 0.00142 |
| 600 | 19.91 | 19.55 | 19.61 | 19.91 | 18.74 | 19.35 | -4.9272 | 7.31E-06 |
| 700 | 19.91 | 19.62 | 19.66 | 19.91 | 18.00 | 19.41 | -3.7358 | 0.00043 |
| 800 | 19.91 | 19.66 | 19.71 | 19.91 | 18.43 | 19.36 | -5.9218 | 1.83E-07 |
| 900 | 19.91 | 19.70 | 19.73 | 19.91 | 18.43 | 19.41 | -4.4924 | 3.41E-05 |
| 1000 | 19.91 | 19.74 | 19.77 | 19.91 | 18.34 | 19.53 | -4.2476 | 7.92E-05 |

**Table 5: Comparison of degree of imbalance obtained by SASOS and CDSOS for right-half distribution dataset**

| Task Size | SASOS | | | CDSOS | | | | |
|---|---|---|---|---|---|---|---|---|
| | max | min | avg | max | min | avg | t-value | p-value |
| 100 | 19.91 | 16.01 | 17.07 | 19.91 | 15.75 | 17.75 | 4.6624 | 1.88E-05 |
| 200 | 19.91 | 18.40 | 18.66 | 19.91 | 15.62 | 18.44 | -1.2319 | 0.22300 |
| 300 | 19.91 | 18.96 | 19.1 | 19.91 | 17.15 | 18.69 | -3.5595 | 0.00075 |
| 400 | 19.91 | 19.21 | 19.34 | 19.91 | 16.51 | 19.04 | -2.2903 | 0.02566 |
| 500 | 19.91 | 19.36 | 19.48 | 19.91 | 16.22 | 19.19 | -2.3254 | 0.02357 |
| 600 | 19.91 | 19.49 | 19.56 | 19.91 | 17.88 | 19.14 | -4.1277 | 0.00012 |
| 700 | 19.91 | 19.56 | 19.63 | 19.91 | 18.37 | 19.37 | -3.8935 | 0.00026 |
| 800 | 19.91 | 19.63 | 19.67 | 19.91 | 18.52 | 19.36 | -4.0421 | 0.00016 |
| 900 | 19.91 | 19.66 | 19.72 | 19.91 | 18.50 | 19.38 | -4.6156 | 2.2E-05 |
| 1000 | 19.91 | 19.71 | 19.76 | 19.91 | 18.67 | 19.57 | -3.3509 | 0.00142 |

**Table 6: Comparison of degree of imbalance obtained by SASOS and CDSOS for uniform distribution dataset**

| Task Size | SASOS | | | CDSOS | | | | |
|---|---|---|---|---|---|---|---|---|
| | max | min | avg | max | min | avg | t-value | p-value |
| 100 | 19.91 | 17.04 | 17.37 | 19.91 | 16.57 | 17.83 | 5.04960 | 4.69E-06 |
| 200 | 19.91 | 18.40 | 18.70 | 19.91 | 17.33 | 18.62 | -0.72619 | 0.47060 |
| 300 | 19.91 | 19.00 | 19.15 | 19.91 | 17.62 | 18.71 | -4.23180 | 8.36E-05 |
| 400 | 19.91 | 19.22 | 19.35 | 19.91 | 17.85 | 18.91 | -4.12710 | 0.00012 |
| 500 | 19.91 | 19.41 | 19.49 | 19.91 | 18.28 | 19.20 | -3.68910 | 0.00050 |
| 600 | 19.91 | 19.50 | 19.58 | 19.91 | 17.74 | 19.19 | -3.88510 | 0.00027 |
| 700 | 19.91 | 19.58 | 19.65 | 19.91 | 16.80 | 19.22 | -3.72020 | 0.00045 |
| 800 | 19.91 | 19.59 | 19.68 | 19.91 | 18.21 | 19.32 | -4.30080 | 6.6E-05 |
| 900 | 19.91 | 19.70 | 19.73 | 19.91 | 18.31 | 19.29 | -5.66680 | 4.8E-07 |
| 1000 | 19.91 | 19.71 | 19.75 | 19.91 | 18.46 | 19.46 | -4.26180 | 7.5E-05 |

The Degree of Imbalance (DI) examines whether the proposed CDSOS task scheduling approach is able to effectively balance tasks on heterogeneous virtual machines, resulting in minimisation of the DI. A smaller value for the DI indicates that the approach is better than the compared SASOS algorithm. Figures 2 - 5 illustrate the performance of the CDSOS approach for task scheduling in terms of the minimum value of the DI achieved for all scheduled tasks compared to the SASOS algorithm. Moreover, the performance shown in the figures indicates that the CDSOS approach for task scheduling achieves a minimum of DI for each task distribution. This is also due to the inclusion of the cuckoo search method, which improves the local search process of SASOS and helps the approach to make effective distribution decisions that do not waste resources due to efficient load balancing.

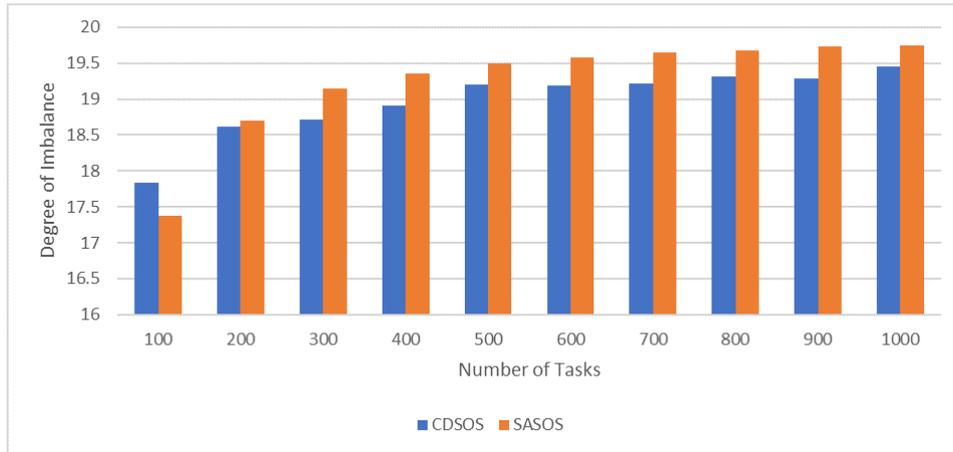

**Figure 2: Degree of Imbalance for Normal Distribution Dataset**

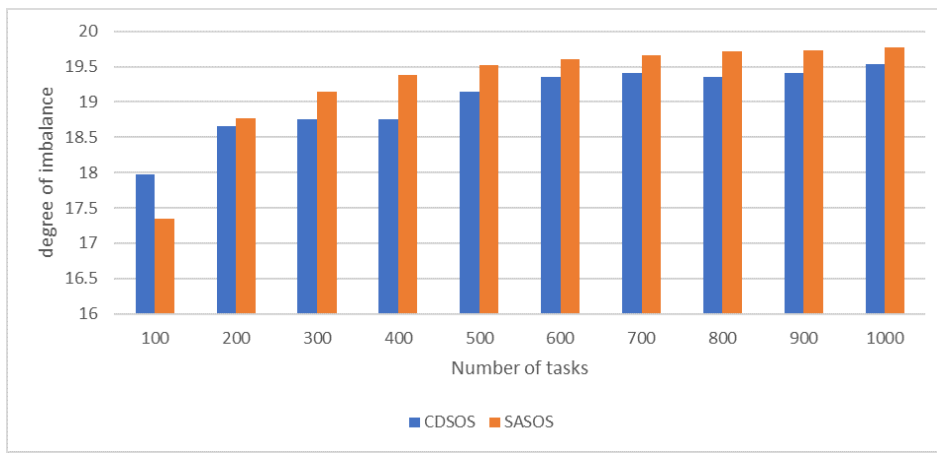

**Figure 3: Degree of Imbalance for Left Half Distribution Dataset**

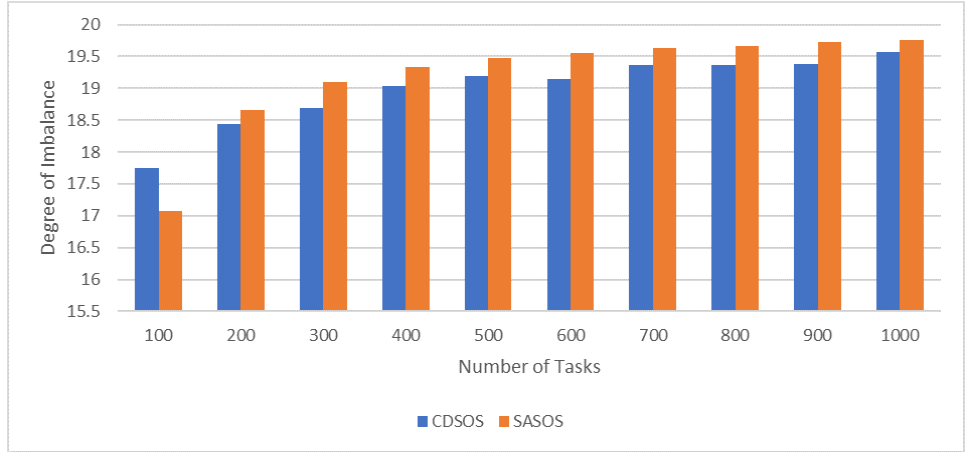

**Figure 4: Degree of Imbalance for Right Half Distribution Dataset**

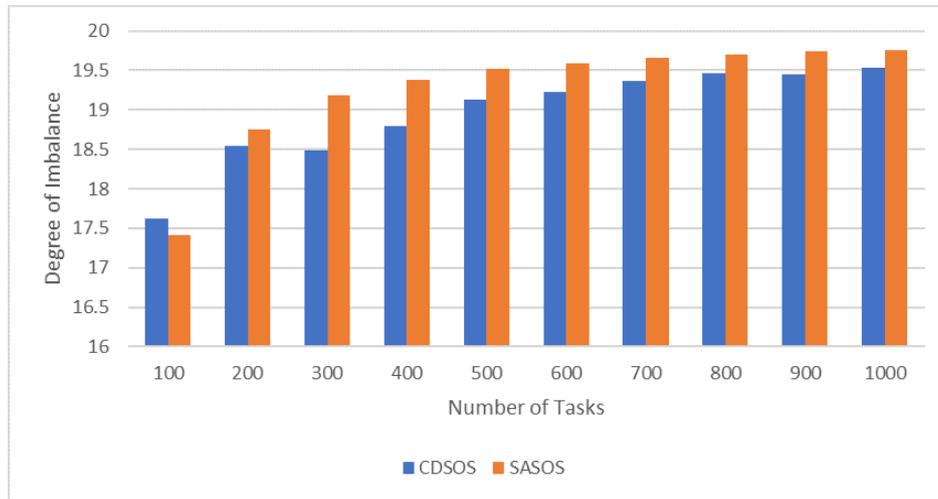

**Figure 5: Degree of Imbalance for Uniform Distribution Dataset**

Moreover, these Figures have shown that the number of virtual machines considered, except for the values for the average degree of imbalance, has no significant impact on the performance of the CDSOS task scheduling strategy. The overall performance has shown that the strategy provides much better performance than the SASOS task scheduling algorithm.

To get a better insight into the performance of the proposed strategy for 100 instance it shown a low performance due to the fact that, it is more effective in handling larger tasks. However, for 200 to 1000 task instances, all Figures show the performance as a function of the degree of imbalance. The CDSOS task scheduling strategy is able to minimise the degree of imbalance compared to the SASOS scheduling algorithm. This shows that the proposed strategy efficiently distributes tasks among the most needed virtual machines, taking into account the computational requirements of each task. Therefore, it significantly increased the convergence rate and contributed to an effective distribution of tasks among virtual machines, reducing the degree of imbalance. In addition, a two-sample t-test was performed to analyse the obtained data. Tables 3 - 6 show the t-values and p-values of all distributions in the data set. In all cases, the p-values are smaller than the alpha value of 0.05, which indicates a significant improvement of the proposed strategy CDSOS over the benchmark SASOS.

## 6. Conclusion

The problem of task scheduling in a cloud computing environment is an NP-hard problem. Cloud performance is mainly affected by task scheduling strategies that are not efficient on cloud resources (i.e., virtual machines) due to higher levels of imbalance. Despite some successes of the SASOS algorithm in addressing the NP-hard problem, it still suffers from a higher degree of imbalance due to the imbalance between the local and global search space. However, similar to other metaheuristics, SASOS cannot guarantee the optimal solution for every NP-hard problem. Consequently, the chance of finding an optimal solution is somewhat difficult with the SASOS, especially when the search space becomes even larger, as in the case of the cloud computing environment. Therefore, this paper has shown that the proposed CDSOS strategy for task scheduling is a novel strategy for task scheduling by incorporating the Levy-Flight part of Cuckoo Search algorithm into the local search of the DSOS (i.e., the mutualism phase), which leads to the optimisation of the degree of imbalance during the task scheduling process. Moreover, the hybrid DSOS gives the proposed strategy the ability to find a better solution in the later phase of the search process, which in turn increases its exploration and exploitation performance. The simulation results using CloudSim simulator have shown that the proposed CDSOS approach has more advantages in optimising the degree of imbalance compared to the benchmark task scheduling algorithm (SASOS).

**Acknowledgement**

The authors wish acknowledge the use of the facilities and services of Universiti Putra Malaysia (UPM), faculty members of Computer Science and Information Technology UPM, and that of Department of Information Technology, Modibbo Adama University Yola Nigeria.

## 7. References

[1] F. Pop and M. Potop-Butucaru, "ARMCO: Advanced topics in resource management for ubiquitous cloud computing: An adaptive approach," *Futur. Gener. Comput. Syst.*, vol. 54, pp. 79–81, 2016.

[2] M. Mezmaz *et al.*, "A parallel bi-objective hybrid metaheuristic for energy-aware scheduling for cloud computing systems," *J. Parallel Distrib. Comput.*, vol. 71, no. 11, pp. 1497–1508, 2011.

[3] Y. Jadeja and K. Modi, "Cloud computing-concepts, architecture and challenges," in *2012 international conference on computing, electronics and electrical technologies (ICCEET)*, 2012, pp. 877–880.

[4] P. Mell, T. Grance, and others, "The NIST definition of cloud computing," 2011.

[5] N. B. Ruparelia, *Cloud computing*. Mit Press, 2023.

[6] M. R. Garey and D. S. Johnson, *Computers and Intractability: A Guide to the Theory of NP-completeness*. Freeman, 1979.

[7] K. Kaur, A. Chhabra, and G. Singh, "Heuristics based genetic algorithm for scheduling static tasks in homogeneous parallel system," *Int. J. Comput. Sci. Secur.*, vol. 4, no. 2, pp. 183–198, 2010.

[8] G. Ming and H. Li, "An Improved Task Scheduling Algorithm based on Max-min for Cloud Computing," *Int. J. Innov. Res. Comput. Commun. Eng. (An ISO Certif. Organ.*, vol. 32972, no. 2, pp. 217–223, 2012.

[9] U. Bhoi and P. N. Ramanuj, "Enhanced max-min task scheduling algorithm in cloud computing," *Int. J. Appl. or Innov. Eng. Manag.*, vol. 2, no. 4, pp. 259–264, 2013.

[10] E. U. Munir, J. Li, and S. Shi, "QoS sufferage heuristic for independent task scheduling in grid," *Inf. Technol. J.*, vol. 6, no. 8, pp. 1166–1170, 2007.

[11] S. M. Abdulhamid, M. S. A. Latiff, and I. Idris, "Tasks scheduling technique using league championship algorithm for makespan minimization in IAAS cloud," *arXiv Prepr. arXiv1510.03173*, 2015.

[12] M. Abdullahi, M. A. Ngadi, and S. M. Abdulhamid, "Symbiotic Organism Search optimization based task scheduling in cloud computing environment," *Futur. Gener. Comput. Syst.*, vol. 56, pp. 640–650, 2016.

[13] S. Sa'ad, A. Muhammed, M. Abdullahi, A. Abdullah, and F. H. Ayob, "An Enhanced Discrete Symbiotic Organism Search Algorithm for Optimal Task Scheduling in the Cloud," *Algorithms*, vol. 14, no. 7, pp. 1–24, 2021.

[14] J. Yu, R. Buyya, and K. Ramamohanarao, "Workflow scheduling algorithms for grid computing," *Metaheuristics Sched. Distrib. Comput. Environ.*, pp. 173–214, 2008.

[15] C. Gogos, C. Valouxis, P. Alefragis, G. Goulas, N. Voros, and E. Housos, "Scheduling independent tasks on heterogeneous processors using heuristics and Column Pricing," *Futur. Gener. Comput. Syst.*, vol. 60, pp. 48–66, 2016.

[16] C. Zhao, S. Zhang, Q. Liu, J. Xie, and J. Hu, "Independent Tasks Scheduling Based on Genetic Algorithm in Cloud Computing," *2009 5th Int. Conf. Wirel. Commun. Netw. Mob. Comput.*, pp. 1–4, 2009.

[17] S. Pandey, L. Wu, S. M. Guru, and R. Buyya, "A particle swarm optimization-based heuristic for scheduling workflow applications in cloud computing environments," *Proc. - Int. Conf. Adv. Inf. Netw. Appl. AINA*, pp. 400–407, 2010.

[18] S. Xue, M. Li, X. Xu, and J. Chen, "An ACO-LB algorithm for task scheduling in the cloud environment," *J. Softw.*, vol. 9, no. 2, pp. 466–473, 2014.

[19] X. Shengjun, Z. Jie, and X. Xiaolong, "An improved algorithm based on ACO for cloud service PDTs scheduling," *Adv. Inf. Sci. Serv. Sci.*, vol. 4, no. 18, 2012.

[20] Z. Xiaolei, M. Congan, and S. Chen, "Task scheduling based on improved bat algorithm under logistics cloud service.," *Appl. Res. Comput. Yingyong Yanjiu*, vol. 32, no. 6, 2015.

[21] S. Raghavan, P. Sarwesh, C. Marimuthu, and K. Chandrasekaran, "Bat algorithm for scheduling workflow applications in cloud," in *2015 International Conference on Electronic Design, Computer Networks & Automated Verification (EDCAV)*, 2015, pp. 139–144.

[22] M.-Y. Cheng, D. Prayogo, and D.-H. Tran, "Optimizing multiple-resources leveling in multiple projects using discrete symbiotic organisms search," *J. Comput. Civ. Eng.*, vol. 30, no. 3, p. 4015036, 2016.

[23] M.-Y. Cheng, D. Prayogo, and D.-H. Tran, "Optimizing Multiple-Resources Leveling in Multiple Projects Using Discrete Symbiotic Organisms Search," *J. Comput. Civ. Eng. 30(3), 04015036.*, vol. 30, no. 3, pp. 287–290, 2016.

[24] R. Eki, F. Y. Vincent, S. Budi, and A. P. Redi, "Symbiotic Organism Search ( SOS ) for Solving the Capacitated Vehicle Routing Problem," vol. 9, no. 5, pp. 873–877, 2015.

[25] D. Gabi, A. S. Ismail, A. Zainal, Z. Zakaria, and A. Abraham, "Orthogonal Taguchi-based cat algorithm for solving task scheduling problem in cloud computing," *Neural Comput. Appl.*, vol. 30, no. 6, pp. 1845–1863, 2018.

[26] D.-H. Tran, M.-Y. Cheng, and D. Prayogo, "A novel Multiple Objective Symbiotic Organisms Search (MOSOS) for time--cost--labor utilization tradeoff problem," *Knowledge-Based Syst.*, vol. 94, pp. 132–145, 2016.

[27] A. A. Zubair *et al.*, "A cloud computing-based modified symbiotic organisms search algorithm (ai) for optimal task scheduling," *Sensors*, vol. 22, no. 4, p. 1674, 2022.

[28] M. Kalra and S. Singh, "A review of metaheuristic scheduling techniques in cloud computing," *Egypt. Informatics J.*, vol. 16, no. 3, pp. 275–295, 2015.

[29] D. Gabi, A. S. Ismail, A. Zainal, Z. Zakaria, and A. Al-Khasawneh, "Hybrid Cat Swarm Optimization and Simulated Annealing for Dynamic Task Scheduling on Cloud Computing Environment," *J. Inf. Commun. Technol.*, vol. 17, no. 3, pp. 435–467, 2018.

[30] M. Abdullahi and M. A. Ngadi, "Hybrid symbiotic organisms search optimization algorithm for scheduling of tasks on cloud computing environment," *PLoS One*, vol. 11, no. 6, pp. 1–29, 2016.

[31] D. Prasad and V. Mukherjee, "A novel symbiotic organisms search algorithm for optimal power flow of power system with FACTS devices," *Eng. Sci. Technol. an Int. J.*, vol. 19, no. 1, pp. 79–89, 2016.

[32] D. Gabi, A. S. Ismail, A. Zainal, Z. Zakaria, A. Abraham, and N. M. Dankolo, "Cloud customers service selection scheme based on improved conventional cat swarm optimization," *Neural Comput. Appl.*, vol. 32, pp. 14817–14838, 2020.

[33] M. H. Nadimi-Shahraki, E. Moeini, S. Taghian, and S. Mirjalili, "Discrete Improved Grey Wolf Optimizer for Community Detection," *J. Bionic Eng.*, pp. 1–28, 2023.

[34] S. M. Abdulhamid, M. S. Abd Latiff, S. H. H. Madni, and M. Abdullahi, "Fault tolerance aware scheduling technique for cloud computing environment using dynamic clustering algorithm," *Neural Comput. Appl.*, vol. 29, no. 1, pp. 279–293, 2018.

[35] D. Gabi, A. S. Ismail, A. Zainal, Z. Zakaria, and A. Abraham, "Orthogonal Taguchi-based cat algorithm for solving task scheduling problem in cloud computing," *Neural Comput. Appl.*, vol. 30, pp. 1845–1863, 2018.

[36] D. Gabi, A. S. Ismail, A. Zainal, Z. Zakaria, and A. Al-Khasawneh, "Hybrid cat swarm optimization and simulated annealing for dynamic task scheduling on cloud computing environment," *J. Inf. Commun. Technol.*, vol. 17, no. 3, pp. 435–467, 2018.

[37] M. Abdullahi, M. A. Ngadi, S. I. Dishing, and S. M. Abdulhamid, "An adaptive symbiotic organisms search for constrained task scheduling in cloud computing," *J. Ambient Intell. Humaniz. Comput.*, vol. 14, no. 7, pp. 8839–8850, 2023.

[38] R. Eki, F. Y. Vincent, S. Budi, and A. A. N. P. Redi, "Symbiotic organism search (SOS) for solving the capacitated vehicle routing problem," *Int. J. Ind. Manuf. Eng.*,

vol. 9, no. 5, pp. 873–877, 2015.

[39] S. H. H. Madni, M. S. Abd Latiff, S. M. Abdulhamid, and J. Ali, "Hybrid gradient descent cuckoo search (HGDCS) algorithm for resource scheduling in IaaS cloud computing environment," *Cluster Comput.*, vol. 22, pp. 301–334, 2019.

[40] M. Wang and W. Zeng, "A comparison of four popular heuristics for task scheduling problem in computational grid," *2010 6th Int. Conf. Wirel. Commun. Netw. Mob. Comput. WiCOM 2010*, pp. 1–4, 2010.

[41] Z. Wu, Z. Ni, L. Gu, and X. Liu, "A revised discrete particle swarm optimization for cloud workflow scheduling," in *2010 international conference on computational intelligence and security*, 2010, pp. 184–188.

[42] Q. Tao, H. Chang, Y. Yi, C. Gu, and Y. Yu, "QoS constrained grid workflow scheduling optimization based on a novel PSO algorithm," in *2009 Eighth International Conference on Grid and Cooperative Computing*, 2009, pp. 153–159.

[43] S. N. Sivanandam and P. Visalakshi, "Scheduling workflow in cloud computing based on hybrid particle swarm algorithm," *Indones. J. Electr. Eng. Comput. Sci.*, vol. 10, no. 7, pp. 1560–1566, 2012.

[44] P.-Y. Yin, S.-S. Yu, P.-P. Wang, and Y.-T. Wang, "Multi-objective task allocation in distributed computing systems by hybrid particle swarm optimization," *Appl. Math. Comput.*, vol. 184, no. 2, pp. 407–420, 2007.

[45] S. N. Sivanandam and P. Visalakshi, "Dynamic task scheduling with load balancing using parallel orthogonal particle swarm optimisation," *Int. J. Bio-Inspired Comput.*, vol. 1, no. 4, pp. 276–286, 2009.

[46] C. Jing and P. Quanke, "Discrete particle swarm optimization algorithm for solving independent task scheduling," *Comput. Eng.*, vol. 34, pp. 214–218, 2008.

[47] P.-Y. Yin, S.-S. Yu, P.-P. Wang, and Y.-T. Wang, "Task allocation for maximizing reliability of a distributed system using hybrid particle swarm optimization," *J. Syst. Softw.*, vol. 80, no. 5, pp. 724–735, 2007.

[48] A. S. Joshi, O. Kulkarni, G. M. Kakandikar, and V. M. Nandedkar, "Cuckoo Search Optimization- A Review," *Mater. Today Proc.*, vol. 4, no. 8, pp. 7262–7269, 2017.

[49] D. T. T. Do, D. Lee, and J. Lee, "Material optimization of functionally graded plates using deep neural network and modified symbiotic organisms search for eigenvalue problems," *Compos. Part B Eng.*, vol. 159, no. June 2018, pp. 300–326, 2018.

[50] D. L. Babu and P. V. Krishna, "Honey bee behavior inspired load balancing of tasks in cloud computing environments," *Appl. Soft Comput. J.*, vol. 13, no. 5, pp. 2292–2303, 2013.

[51] R. N. Calheiros, R. Ranjan, A. Beloglazov, C. A. F. De Rose, and R. Buyya, "CloudSim: a toolkit for modeling and simulation of cloud computing environments and evaluation of resource provisioning algorithms," *Softw. Pract. Exp.*, vol. 41, no. 1, pp. 23–50, 2011.

[52] S. M. S. M. Abdulhamid *et al.*, "An adaptive symbiotic organisms search for constrained task scheduling in cloud computing," *arXiv Prepr. arXiv1510.03173*, vol. 22, no. 12, pp. 8839–8850, 2022.

[53] M. Abdullahi, M. A. Ngadi, S. I. Dishing, S. M. Abdulhamid, and B. I. eel Ahmad, "An efficient symbiotic organisms search algorithm with chaotic optimization strategy for multi-objective task scheduling problems in cloud computing environment," *J. Netw. Comput. Appl.*, vol. 133, pp. 60–74, 2019.

[54] S. H. H. Madni, M. S. A. Latiff, Y. Coulibaly, and S. M. Abdulhamid, "Recent advancements in resource allocation techniques for cloud computing environment: a systematic review," *Cluster Comput.*, vol. 20, no. 3, pp. 2489–2533, 2017.

[55] S. 'i M. Abdulhamid, "Task Scheduling Techniques with Fault Tolerance Awareness for Cloud Computing Environment," 2016.

[56] M. Abdullahi, M. A. Ngadi, and S. I. Dishing, "Chaotic Symbiotic Organisms Search for Task Scheduling Optimization on Cloud Computing Environment," pp. 1–4, 2017.